\documentclass[conference,letterpaper]{IEEEtran}
\IEEEoverridecommandlockouts
\usepackage[utf8]{inputenc}
\usepackage[T1]{fontenc}
\usepackage{url}
\usepackage{ifthen}
\usepackage{cite}
\usepackage[cmex10]{amsmath}  
\usepackage{amssymb}
\usepackage{mathrsfs}
\usepackage{graphicx}
\usepackage{bm}
\usepackage{color}
\newtheorem{proposition}{\underline{Proposition}}

\newtheorem{remark}{\underline{Remark}}
\begin{document}
\title{Integrated Sensing and Communication Exploiting Prior Information: How Many Sensing Beams are Needed? }
\author{\IEEEauthorblockN{Chan Xu and Shuowen Zhang}
	\IEEEauthorblockA{Department of Electrical and Electronic Engineering,
		The Hong Kong Polytechnic University\\
		E-mail: \{chan.xu, shuowen.zhang\}@polyu.edu.hk}
}
\maketitle

\begin{abstract}
	This paper studies an integrated sensing and communication (ISAC) system where a multi-antenna base station (BS) aims to communicate with a single-antenna user in the downlink and sense the \emph{unknown} and \emph{random} angle parameter of a target via exploiting its prior distribution information. We consider a general transmit beamforming structure where the BS sends one communication beam and potentially one or multiple dedicated sensing beam(s). Firstly, motivated by the periodic feature of the angle parameter, we derive the \emph{periodic posterior Cram\'{e}r-Rao bound (PCRB)} for quantifying a lower bound of the mean-cyclic error (MCE), which is more accurate than the conventional PCRB for bounding the mean-squared error (MSE). Then, note that more sensing beams enable higher flexibility in enhancing the sensing performance, while also generating extra interference to the communication user. To resolve this trade-off, we formulate the transmit beamforming optimization problem to minimize the periodic PCRB subject to a communication rate requirement for the user. Despite the non-convexity of this problem, we derive the \emph{optimal solution} by leveraging the semi-definite relaxation (SDR) technique and Lagrange duality theory. Moreover, we analytically prove that \emph{at most one} dedicated sensing beam is needed. Numerical results validate our analysis and the advantage of having a dedicated sensing beam.
\end{abstract}
\vspace{-1mm}
\section{Introduction}
\vspace{-1mm}
Integrated sensing and communication (ISAC) is a promising technology for providing high-rate communication and high-precision sensing services in sixth-generation (6G) wireless networks \cite{ISAC_survey2,ISAC_survey1}. By properly utilizing the spatial dimensions provided by multi-antenna base stations (BSs), ISAC can be realized via sending beamformed signals from the BS, which will be both decoded at the communication users and received back at the BS via target reflection. To accomplish both functions effectively under a shared amount of spatial and power resources at the BS, it is of paramount importance to study the beamforming design at the BS in an ISAC system.

Along this research direction, most existing works considered the scenario where the target's parameter to be sensed is \emph{deterministic}, and the beamforming design is based on the sensing performance corresponding to a \emph{given (known) parameter} (see, e.g., \cite{jointDesign_approximation,jointDesign_MI,jointDesign_CRB1,jointDesign_CRB2,Xujie2023,Hua2024}). However, in practice, the parameter to be sensed may be \emph{unknown} and \emph{random}, for which only the probability density function (PDF) can be known \emph{a priori} via historic data and statistical analysis \cite{ISIT,JSAC_MIMO,HouKaiyue1,HouKaiyue2}. By exploiting such prior information, a so-called \emph{posterior Cram\'{e}r-Rao bound (PCRB)}, or Bayesian Cram\'{e}r-Rao bound (BCRB), can be derived to quantify a lower bound of the mean-squared error (MSE) in sensing random parameters \cite{VanTrees}. Compared to the conventional Cram\'{e}r-Rao bound (CRB) which quantifies an MSE lower bound for sensing deterministic parameters, PCRB generally takes a more complicated form that is dependent on the parameter's PDF, based on which study of the transmit beamforming design is still in its infancy \cite{ISIT,JSAC_MIMO,HouKaiyue1,HouKaiyue2,Yuan_PCRB,Yu,Caire}.

Note that when the parameter to be sensed is unknown and random, the transmit beamforming design generally needs to cater to multiple possible values of the parameter (e.g., a continuous range of possible target locations) based on its PDF. To satisfy this high demand, \emph{extra dedicated sensing beams} may be needed at the transmitter to enhance the design flexibility, especially when the number of available communication beams is limited. However, extra sensing beams will also generate interference to the communication users, thus leading to a trade-off between communication and sensing. How many dedicated sensing beams are needed in an ISAC system with unknown and random parameters and how to design the corresponding transmit beamforming still remain open problems, which motivate the study in this paper.

\begin{figure}[t]
	\centering
	\includegraphics[width=2.5in]{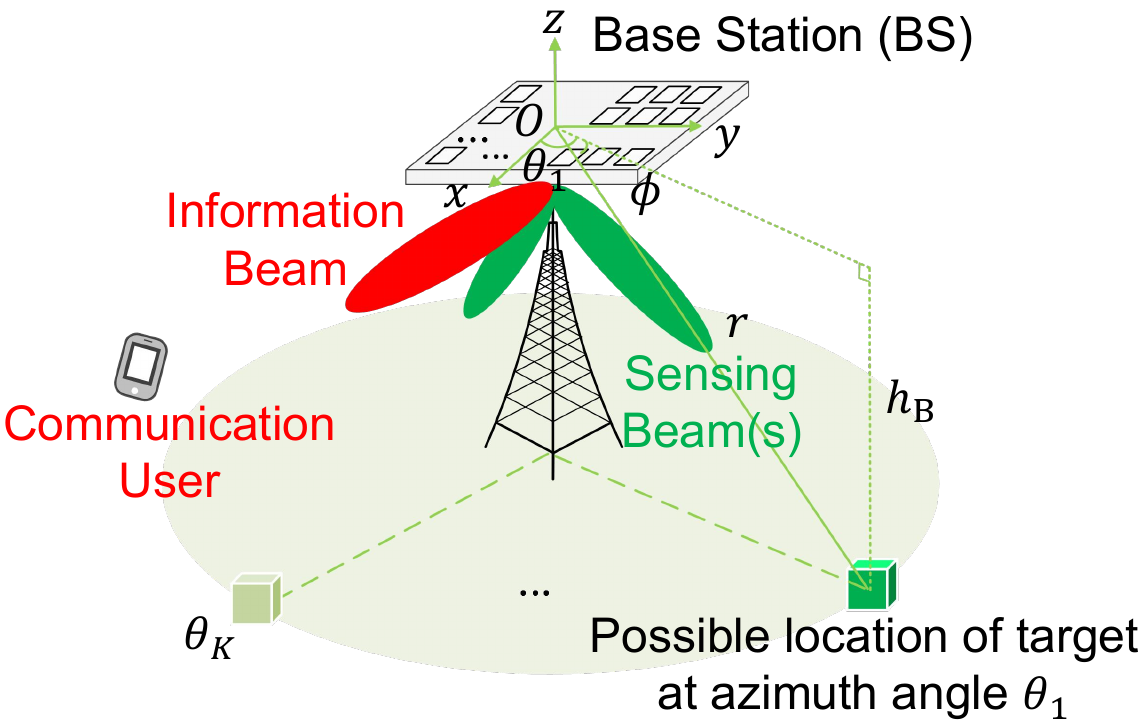}
	\vspace{-4mm}
	\caption{Illustration of an ISAC system with dedicated sensing beam(s) for sensing the unknown and random angle parameter of a target.}\label{fig_sys}
	\vspace{-8mm}	
\end{figure}

This paper aims to address the above problems in an ISAC system where a multi-antenna BS communicates with a single-antenna user in the downlink and senses the \emph{unknown} and \emph{random} angle parameter of a target via its reflected echo signals at the BS receive antennas. The PDF of the angle parameter is known \emph{a priori} for exploitation. We first characterize the sensing performance via deriving the \emph{periodic PCRB} of the sensing mean-cyclic error (MCE), which is tailored for \emph{periodic} parameters (e.g., angle) and is more accurate compared to the conventional PCRB of the MSE. Then, under a transmit beamforming structure with one communication beam and potentially one or multiple dedicated sensing beam(s), we formulate the beamforming optimization problem to minimize the periodic PCRB subject to a rate requirement at the communication user, which is a non-convex problem. By adopting the semi-definite relaxation (SDR) technique and proving the relaxation is always tight, we derive the \emph{optimal solution} and analytically prove that \emph{at most one} sensing beam is needed. Our results are validated by numerical results.
\vspace{-2mm}
\section{System Model}
\vspace{-1mm}
Consider an ISAC system where a BS equipped with $N_t\geq 1$ transmit antennas and $N_r\geq 1$ co-located receive antennas delivers information to a single-antenna communication user in the downlink and estimates the \emph{unknown} and \emph{random} location parameter of a ground target via its reflected echo signals. The antennas at the BS are deployed under a uniform planar array (UPA) configuration. Specifically, we consider a three-dimensional (3D) spherical coordinate system with a reference point at the BS being the origin, as illustrated in Fig. \ref{fig_sys}. Let $h_{\mathrm{B}}\geq 0$ denote the height of the BS antennas in meters (m). We assume the target is located at a known distance (range) of $r$ m from the BS, which leads to a known elevation angle of $\phi=\arcsin\left(\frac{-h_{\mathrm{B}}}{r}\right)$ \cite{ISIT}. Thus, the only \emph{unknown} and \emph{random} parameter is the azimuth angle denoted by $\theta\in[-\pi,\pi)$. We assume the PDF of $\theta$ denoted by $p_\Theta(\theta)$ is known \emph{a priori}.\footnote{Such PDF can be obtained via historic data, target location pattern, etc.}

We aim to study the transmit beamforming design within $L\geq 1$ symbol intervals, during which both the target's location and the communication channel between the BS and the user remain unchanged. Let $\bm{h}^H \in \mathbb{C}^{1\times N_t}$ denote the BS-user channel, which is assumed to be known perfectly at both the BS and the user. Note that for the case without the sensing function, the optimal transmit signal design is to send one communication beam to the single-antenna user using maximum ratio transmission (MRT) beamforming. On the other hand, to accurately sense the target's location, the transmit signal needs to focus high power towards the locations with high probabilities for striking strong echo signals. Since the user's location and the target's possible locations may be significantly different, using one dual-functional beam for both communication and sensing may be strictly suboptimal.\footnote{For instance, when the BS-user channel is orthogonal to the channel from the BS to the most-probable target's location, a dual-functional beam may not be able to achieve high rate and strike strong echo signals at the same time.}

Motivated by the above, we consider a general design method where the transmit signal is the superposition of both a \emph{communication beam} and one or multiple \emph{dedicated sensing beams}.\footnote{Note that we consider a single communication beam for ease of detection at the single-antenna user.} Let $c_l\sim\mathcal{CN}(0,1)$ denote the information symbol for the communication user in each $l$-th symbol interval, and $\bm{\upsilon}_l\in\mathbb{C}^{N_t\times 1}$ denote the signals dedicated for target sensing in the $l$-th symbol interval, in which each element has unit average power and is independent of other elements and $c_l$, with $\mathbb{E}[\bm{\upsilon}_l\bm{\upsilon}_l^H]=\bm{I}_{N_t}$. Let $\bm{w}\in\mathbb{C}^{N_t\times 1}$ denote the information beamforming vector, and $\bm{S}=[\bm{s}_1,...,\bm{s}_{N_t}] \in \mathbb{C}^{N_t\times N_t}$ denote the sensing beamforming matrix with $\bm{s}_i \in \mathbb{C}^{N_t\times 1}$ denoting the beamforming vector for the $i$-th sensing signal. The baseband equivalent transmitted signal vector in the $l$-th symbol interval is thus given by $
\bm{x}_l=\bm{w}c_l+\bm{S}\bm{\upsilon}_l,\ l=1,...,L$.
The transmit covariance matrix is given by $\bm{R}_X=\mathbb{E}[\bm{x}_l\bm{x}_l^H]=\bm{w}\bm{w}^H +\bm{S}\bm{S}^H$. Let $P$ denote the transmit power constraint at the BS transmit antennas, which yields $\|\bm{w}\|^2+\mathrm{tr}(\bm{SS}^H)\leq P$.
\vspace{-2mm}
\subsection{Communication Model and Performance}
\vspace{-1mm}
At the communication user receiver, the received signal in each $l$-th symbol interval is given by $
y_l^{\mathrm{C}}=\bm{h}^H\bm{w}{c}_l+\bm{h}^H\bm{S}\bm{\upsilon}_l+n_l^{\mathrm{C}},\ l=1,...,L$,
where $n_l^{\mathrm{C}}\sim\mathcal{CN}(0,\sigma_{\mathrm{C}}^2)$ denotes the circularly symmetric complex Gaussian (CSCG) noise in the $l$-th symbol interval, with $\sigma_{\mathrm{C}}^2$ denoting the average noise power. We consider a challenging scenario where the sensing signals $\bm{\upsilon}_l$'s are not known at the user receiver, and follow the CSCG distribution corresponding to the worst-case interference for the user. The achievable rate is thus given by
\vspace{-1.5mm}
\begin{equation}\label{rate}
	R=\log_2\left(1+\frac{|\bm{h}^H\bm{w}|^2}{\|\bm{h}^H\bm{S}\|^2+\sigma_{\mathrm{C}}^2}\right)  	
\vspace{-1.5mm}
\end{equation}
in bits per second per Hertz (bps/Hz).
\vspace{-2mm}
\subsection{Sensing Model}\label{sec_sensing}
\vspace{-1mm}
On the other hand, we consider that the target is located within the line-of-sight (LoS) range of the BS. Let $\bm{a}(\phi,\theta)\in\mathbb{C}^{N_t\times 1}$ and $\bm{b}(\phi,\theta)\in \mathbb{C}^{N_r\times 1}$ denote the steering vectors of the BS transmit and receive antenna arrays, respectively. Specifically, $a_n(\phi,\theta)\!=\!e^{j\Delta[(T_x\!-\!2\lfloor\frac{n-1}{T_y}\rfloor\!-\!1)\cos\theta\!+\!(T_y\!-\!2((n-1)\!\!\!\mod\!T_y)\!-\!1)\sin\theta]}$ and $b_m(\phi,\theta)$ $\!=\!e^{j\Delta[(R_x\!-\!2\lfloor\frac{m-1}{R_y}\rfloor\!-\!1)\cos\!\theta\!+\!(R_y\!-\!2((m-1)\!\!\!\mod R_y)\!-\!1)\sin\theta]}$, where $\Delta=\frac{-\pi d\cos\phi}{\lambda}$ with $d$ denoting the antenna spacing and $\lambda$ denoting the wavelength; $T_x$, $R_x$ and $T_y$, $R_y$ denote the numbers of rows and columns at the transmit/receive arrays, respectively, with $N_t\!=\!T_xT_y$ and $N_r\!=\!R_xR_y$. Let $\psi\in \mathbb{C}$ denote the radar cross-section (RCS) coefficient, which is assumed to be an unknown and deterministic parameter. The overall channel from the BS transmitter to the BS receiver via target reflection is thus given by $ \bm{G}(\theta)\!=\!\frac{\sqrt{\beta_0}}{r}\bm{b}(\phi,\theta)\psi\frac{\sqrt{\beta_0}}{r}\bm{a}^H(\phi,\theta)\!\overset{\Delta}{=}\!\alpha \bm{b}(\phi,\theta)\bm{a}^H(\phi,\theta)$, where $\beta_0$ denotes the reference channel power at reference distance $1$ m, and $\alpha\overset{\Delta}{=}\frac{\beta_0}{r^2}\psi=\alpha_{\mathrm{R}}+j\alpha_{\mathrm{I}}$ characterizes the round-trip path loss and target reflection. Note that $\alpha$ is an \emph{unknown} and \emph{deterministic} parameter. The received echo signal at the BS receive antennas in the $l$-th symbol interval is given by $
\bm{y}_l^{\mathrm{S}}=\bm{G}(\theta)\bm{x}_l+{\bm{n}_l^{\mathrm{S}}},\ l=1,...,L$,
where $ \bm{n}_l^{\mathrm{S}} \sim\mathcal{CN}(\bm{0},\sigma_{\mathrm{S}}^2\bm{I}_{N_r})$  denotes the CSCG noise vector over the BS receive antennas in the $l$-th symbol interval, with $\sigma_{\mathrm{S}}^2$ denoting the average noise power. The collection of received echo signals over $L$ symbol intervals is thus given by $
\bm{Y}=[{\bm{y}_1^{\mathrm{S}}},...,{\bm{y}_L^{\mathrm{S}}} ]=\bm{G}(\theta)[ \bm{x}_1 ,...,\bm{x}_L ]+[{\bm{n}_1^{\mathrm{S}}},...,{\bm{n}_L^{\mathrm{S}}} ]$.
It is worth noting that both $c_l$'s and $\bm{\upsilon}_l$'s in $\bm{x}_l$'s are known at the BS receiver since they are generated from the co-located BS transmitter. Hence, \emph{both} the communication beam and the sensing beams can be used for sensing. Based on this, the BS can estimate the unknown and random parameter $\theta$ jointly with the unknown and deterministic parameter $\alpha$ from $\bm{Y}$.

Unlike the achievable communication rate in (\ref{rate}), derivation of the performance for sensing $\theta$ via exploiting its prior PDF is not straightforward, particularly since $\theta$ is an angle parameter with periodic feature. In the next section, we propose to characterize the sensing performance in terms of the transmit beamforming vectors by deriving a so-called \emph{periodic PCRB} for the \emph{MCE}, which is more suitable for periodic parameters compared to conventional bounds for the MSE.
\vspace{-2mm}
\section{Sensing Performance Characterization Based on Periodic PCRB}
\vspace{-1.5mm}
Let $\bm{\zeta}=[\theta,\alpha_{\mathrm{R}},\alpha_{\mathrm{I}}]^T$ denote the collection of all the unknown parameters to be estimated (or sensed) by jointly exploiting the observation $\bm{Y}$ and the prior PDF $p_\Theta(\theta)$. Note that due to the \emph{periodic} feature of the angle parameter $\theta$ and its effect on the overall reflected channel $\bm{G}(\theta)$, quantifying the sensing performance via the direct difference between the actual and estimated values may not be appropriate. For instance, if the actual value of $\theta$ is $\pi-\epsilon_1$ and the estimated value is $-\pi+\epsilon_2$, where $\epsilon_1$ and $\epsilon_2$ are both small positive values, the direct sensing error is $-2\pi+(\epsilon_1+\epsilon_2)$. However, the actual \emph{periodic or cyclic error} is $\epsilon_1+\epsilon_2$, which is much less significant and can better reflect the sensing accuracy since $\theta\in[-\pi,\pi)$ has a $2\pi$-periodic feature. Motivated by this, we propose to quantify a lower bound of the MCE as follows.

Let $\hat{\theta}$ denote an estimator of $\theta$. The MCE for measuring the error of estimating $\theta$ is given by $
\mathrm{MCE}(\hat{\theta})=2-2\mathbb{E}_{\bm{Y},\bm{\zeta}}[\cos(\hat{\theta}-\theta)]$ \cite{MCE1,periodic_PCRB}. Note that the direct estimation error of $\theta$ without exploiting the periodic feature of $\theta$ is given by $\epsilon=\hat{\theta}-\theta$, and the MSE is given by $\mathrm{MSE}(\hat{\theta}  )=\mathbb{E}_{\bm{Y},\bm{\zeta}}[|\hat{\theta}-\theta|^2]$. Notice that the MCE is always no larger than the MSE due to $2-2\cos\epsilon\leq\epsilon^2,\forall \epsilon \in \mathbb{R}$, since the ``dummy'' sensing errors can be excluded.

Define $\bm{F}$ as the periodic posterior Fisher information matrix (FIM) for $\bm{\zeta}$, based on which the periodic PCRB of the MCE for the desired sensing parameter $\theta$ is given by $\mathrm{PCRB}^{\mathrm{P}}_{\theta}\!=\! 2-2\left(1\!+\! [\bm{F}^{-1}]_{1,1}\right)^{-\frac{1}{2}}$ \cite{periodic_PCRB}, which is further derived in the following as a function of the PDF of $\theta$, $p_{\Theta}(\theta)$.

\begin{proposition}\label{prop_PCRB}
	The periodic PCRB for the MCE in estimating $\theta$ is given by
	\begin{align}\label{PCRB}
&\mathrm{PCRB}^{\mathrm{P}}_{\theta}=2-\frac{2}{(1+ [\bm{F}^{-1}]_{1,1})^{\frac{1}{2}}}  \\[-0.5mm]
		=&2-\frac{2}{\left( \!\!1\!+\!\frac{1}{\mathbb{E}_\theta\left[\!\left(\frac{\partial \ln(\hat{p}_\Theta(\theta))}{\partial \theta}\right)^2\!\right]+ \frac{2|\alpha|^2L}{\sigma_{\mathrm{S}}^2}\left( \!\mathrm{tr}\left(\bm{A}_1\!\bm{R}_X\right) - \frac{\left| \mathrm{tr}(\!\bm{A}_2\!\bm{R}_X\!)\right|^2}{ \mathrm{tr}(\!\bm{A}_3\!\bm{R}_X\!)}\!\right)} \!\!\right)^{\!\frac{1}{2}}},\!\!\!\nonumber
	\end{align}
	where $\hat{p}_\Theta(\theta) =  {p}_\Theta(\theta-2\pi\lfloor\frac{\theta+\pi}{2\pi}\rfloor),\theta\in \mathbb{R}$, $\bm{A}_1\!\!=\!\! \int_{-\pi}^\pi \!\left(\!\|\dot{\bm{b}}(\phi,\!\theta)\|^2\bm{a}(\phi,\!\theta)\bm{a}^H\!(\phi,\!\theta)\!+\!N_r\dot{\bm{a}}(\phi,\!\theta)\dot{\bm{a}}^H\!(\phi,\!\theta)\!\right) \!p_\Theta(\theta)d\theta $, $\bm{A}_2\!=\!N_r\int_{-\pi}^\pi\dot{\bm{a}}(\phi,\theta)\bm{a}^H(\phi,\theta) p_\Theta(\theta)d\theta $, and $\bm{A}_3=N_r\int_{-\pi}^\pi\bm{a}(\phi,\theta)\bm{a}^H(\phi,\theta) p_\Theta(\theta)d\theta$, with $\dot{\bm{a}}(\phi,\theta)$, $\dot{\bm{b}}(\phi,\theta)$ denoting the derivatives of ${\bm{a}}(\phi,\theta)$, ${\bm{b}}(\phi,\theta)$ with respect to $\theta$.
\end{proposition}
\begin{IEEEproof}
	Please refer to Appendix A.
\end{IEEEproof}

\begin{remark}
It is worth noting that the periodic PCRB derived in Proposition \ref{prop_PCRB} is applicable to any
PDF in $[-\pi,\pi)$. As an example, we provide a more explicit form of it under a practical and typical PDF. Motivated by practical scenarios where the target's azimuth angles are typically concentrated around several nominal angles, we consider a \emph{von-Mises mixture PDF}, which is the weighted summation of $K$ von-Mises PDFs, each being similar to a Gaussian distribution on a circle \cite{directional}. For each $k$-th von-Mises PDF, $\theta_k\in[-\pi,\pi)$ denotes the mean angle, $\kappa_k>0$ denotes the concentration, and $p_k\in [0,1]$ denotes the weight. The PDF of $\theta$ is thus given by $p_\Theta(\theta)=\sum_{k=1}^{K}p_k \frac{e^{\kappa_k\cos(\theta-\theta_k)}}{2\pi I_0(\kappa_k)},\ \theta\in[-\pi,\pi)$, where $I_n(\kappa_k)=\frac{1}{2\pi}\int_{-\pi}^\pi e^{\kappa_k\cos t}\cos(nt)dt$ is the modified Bessel function of order $n$.\footnote{Note that with different parameters, the considered von-Mises mixture PDF can characterize various practical scenarios. For instance, if $\kappa_k\rightarrow 0,\forall k $, the PDF will approach a uniform PDF; if $\kappa_k\rightarrow \infty,\forall k$, the PDF will approach a probability mass function (PMF) with $K$ possible angles at $\theta_k$'s.} In this case, due to the $2\pi$-periodicity of the cosine function, $\hat{p}_\Theta(\theta) =  {p}_\Theta(\theta-2\pi\lfloor\frac{\theta+\pi}{2\pi}\rfloor)=\sum_{k=1}^{K}p_k \frac{e^{\kappa_k\cos(\theta-\theta_k-2\pi\lfloor\frac{\theta+\pi}{2\pi}\rfloor)}}{2\pi I_0(\kappa_k)}=\sum_{k=1}^{K}p_k \frac{e^{\kappa_k\cos(\theta-\theta_k)}}{2\pi I_0(\kappa_k)}$. Then,  $\mathbb{E}_\theta\big[\big(\frac{\partial \ln(\hat{p}_\Theta(\theta))}{\partial \theta}\big)^2\big]=\sum\limits_{k=1}^{K}p_k\int_{-\pi}^{\pi}  \hat{f}_k(\theta)\kappa_{k}^2\sin^2(\theta-\theta_{k})d\theta-\rho =\sum\limits_{k=1}^{K} \frac{p_k\kappa_kI_1(\kappa_k)}{I_0(\kappa_k)} -\rho$, where $\hat{f}_{k}(\theta)\!\overset{\Delta}{=}\!\frac{e^{\kappa_k\cos(\theta-\theta_k)}}{2\pi I_0(\kappa_k)}$ and $\rho\overset{\Delta}{=}\int_{-\pi}^{\pi}  \sum\limits_{k_1=1}^{K}\sum\limits_{k_2=1 }^{K}p_{k_1}p_{k_2}\hat{f}_{k_1}(\theta)\hat{f}_{k_2}(\theta)( \kappa_{k_1}\sin(\theta-\theta_{k_1})-\kappa_{k_2}\sin(\theta-\theta_{k_2})  )^2 /\big(2\sum\limits_{k=1}^{K}p_k\hat{f}_k(\theta)\big)   d\theta$. Hence, $[\bm{F}^{-1}]_{1,1}=1\Big/\Big(\sum\limits_{k=1}^{K}\frac{p_k\kappa_kI_1(\kappa_k)}{I_0(\kappa_k)}-\rho	+\frac{2|\alpha|^2L}{\sigma_\mathrm{S}^2}\Big( \mathrm{tr}(\bm{A}_1\bm{R}_X) - \frac{| \mathrm{tr}(\bm{A}_2\bm{R}_X)|^2}{ \mathrm{tr}(\bm{A}_3\bm{R}_X)}\Big)\Big)$.
\end{remark}

Notice that $\mathrm{PCRB}^{\mathrm{P}}_{\theta}$ for the MCE of estimating $\theta$ is determined by the transmit covariance matrix $\bm{R}_X=\bm{ww}^H+\bm{SS}^H$ and consequently $\bm{w}$ and $\bm{S}$. Moreover, the communication beam $\bm{w}$ and each sensing beam in $\bm{S}$ play the same role in the periodic PCRB. Having more sensing beams will add more flexibility in focusing the transmit signals to more highly-probable angles for lowering $\mathrm{PCRB}^{\mathrm{P}}_{\theta}$. On the other hand, the dedicated sensing beams will cause extra interference to the communication user, thus limiting the communication rate. In the following, we study the optimization of $\bm{w}$ and $\bm{S}$ to achieve an optimal trade-off between communication and sensing.

\vspace{-1mm}
\section{Problem Formulation}
\vspace{-1mm}
We aim to optimize the information beamforming vector $\bm{w}$ and sensing beamforming matrix $\bm{S}$ to minimize the sensing periodic PCRB subject to a communication rate target denoted by $\bar{R}>0$ bps/Hz. The optimization problem is formulated as
\begin{align}
	\mbox{(P1)}  \quad \mathop{\mathrm{min}}_{\bm{w} ,\bm{S}  }\quad&  \mathrm{PCRB}^{\mathrm{P}}_{\theta}\\
	\mathrm{s.t.}\quad & \log_2\left(1+\frac{|\bm{h}^H \bm{w}|^2 }{\|\bm{h}^H \bm{S}\|^2+\sigma_{\mathrm{C}}^2}\right)\geq\bar{R}  \label{P1_C1}  \\
	& \|\bm{w}\|^2+\mathrm{tr}(\bm{SS}^H)\leq P.\label{P1c2}
\end{align}
Note that the objective function of Problem (P1) involves a complicated fractional structure, and can be shown to be non-convex. Moreover, the constraint in (\ref{P1_C1}) is also non-convex. Therefore, Problem (P1) is a non-convex problem. In the following, we obtain the optimal solution to (P1) via SDR and unveil the optimal number of sensing beams.

\section{Optimal Solution to Problem (P1)}\label{sec_solution}
\subsection{Problem Transformation}
Note that $\mathrm{PCRB}^{\mathrm{P}}_{\theta}$ decreases as $g(\bm{w},\bm{S})\triangleq\mathrm{tr}(\bm{A}_{1}(\bm{w}\bm{w}^H+\bm{SS}^H))-\frac{|\mathrm{tr}(\bm{A}_2(\bm{w}\bm{w}^H+\bm{S\!S}^H))|^{2}}{ \mathrm{tr}(\bm{A}_3 (\bm{w}\bm{w}^H+\bm{SS}^H))}$ increases. By further defining $\gamma=2^{\bar{R}}-1$ and introducing an auxiliary variable $t$, Problem (P1) is equivalent to the following problem:
\begin{align} \label{P2}
	\mbox{(P2)}\quad \mathop{\mathrm{max}}_{t,\bm{w},\bm{S}:(\ref{P1c2})}\quad  &t\\[-1mm]
	\mathrm{s.t.} \quad  &g(\bm{w},\bm{S})\geq t\label{P2c1}\\[-1mm]
	& |\bm{h}^H \bm{w}|^2\geq\gamma (\|\bm{h}^H\bm{S}\|^2+\sigma_{\mathrm{C}}^2).  \label{P2_C1}
\end{align}

Define $\bm{R}_{\mathrm{C}}\triangleq\bm{w}\bm{w}^H $ with $\mathrm{rank}(\bm{R}_{\mathrm{C}})=1$ and $\bm{R}_{\mathrm{S}}\triangleq\bm{SS}^H$. Based on Schur complement \cite{schur}, (\ref{P2c1}) is equivalent to
\vspace{-1.5mm}
\begin{align}
	&\bm{B}(t,\bm{R}_{\mathrm{C}},\bm{R}_{\mathrm{S}})\nonumber\\[-0.5mm]\overset{\Delta}{=}&\left[\begin{array}{ll}
		\mathrm{tr}(\bm{A}_1(\bm{R}_{\mathrm{C}}+\bm{R}_{\mathrm{S}}))-t &\mathrm{tr}(\bm{A}_2(\bm{R}_{\mathrm{C}}+\bm{R}_{\mathrm{S}})) \\[-0.1mm]
		\mathrm{tr}(\bm{A}_2^H(\bm{R}_{\mathrm{C}}+\bm{R}_{\mathrm{S}})) &\mathrm{tr}(\bm{A}_3(\bm{R}_{\mathrm{C}}+\bm{R}_{\mathrm{S}}))
	\end{array}\right]  \!\succeq \! \bm{0}.\!\!\!
\end{align}
By further applying the SDR technique, Problem (P2) and Problem (P1) are equivalent to the following problem with an additional constraint of $\mathrm{rank}(\bm{R}_{\mathrm{C}})=1$:
\begin{align} \label{P2-R}
	\mbox{(P2-R)} \quad  \mathop{\mathrm{max}}_{t,\bm{R}_{\mathrm{C}},\bm{R}_{\mathrm{S}}} \quad & t
	\\[-1mm]
	\mathrm{s.t.} \quad  & \bm{B}\left(t,\bm{R}_{\mathrm{C}},\bm{R}_{\mathrm{S}}\right) \succeq \bm{0}  \label{P2R_C1}\\[-1mm]
	&  \bm{h}^H \bm{R}_{\mathrm{C}}\bm{h} \geq  {\gamma} \left(\sigma_{\mathrm{C}}^2 + \bm{h}^H \bm{R}_{\mathrm{S}}\bm{h}\right) \label{P2R_C2}\\[-1mm]
	& \mathrm{tr}(\bm{R}_{\mathrm{C}}  +  \bm{R}_{\mathrm{S}})\leq P  \label{P2R_C3}\\[-1mm]
	&  \bm{R}_{\mathrm{C}} \succeq \bm{0}\label{P2R_C4}\\[-1mm]
	& \bm{R}_{\mathrm{S}} \succeq \bm{0}. \label{P2R_C5}
\end{align}
Note that Problem (P2-R) is a convex optimization problem, for which the optimal solution can be obtained via the interior-point method \cite{CVX} or existing software, e.g., CVX \cite{cvxtool}. Note that if the optimal solution to (P2-R) denoted by $(t^\star,\bm{R}^\star_{\mathrm{C}},\bm{R}^\star_{\mathrm{S}})$ yields $\mathrm{rank}(\bm{R}^\star_{\mathrm{C}})=1$, the SDR relaxation from (P2) to (P2-R) is \emph{tight}, and an optimal solution to (P2) and (P1) can be obtained via $\bm{R}^\star_{\mathrm{C}}=\bm{w}^\star\bm{w}^{\star^H}$ and $\bm{R}^\star_{\mathrm{S}}=\bm{S}^\star\bm{S}^{\star^H}$. In the following, we analytically prove that the SDR relaxation is always tight, and further obtain an optimal solution to (P1).
\vspace{-2mm}
\subsection{Properties of the Optimal Solution to Problem (P2-R)}
\vspace{-1mm}
We employ the Lagrange duality theory \cite{CVX} to analyze the properties of the optimal solution to (P2-R). First, we introduce $\bm{Z}_B=[z_1,z_2;z_2^*,z_3]$, $\mu_R$, $\mu_P$, $\bm{Z}_{\mathrm{C}}$, and $\bm{Z}_{\mathrm{S}}$ as the dual variables associated with the constraints in (\ref{P2R_C1})-(\ref{P2R_C5}), respectively. The Lagrangian of (P2-R) is given by
\begin{align}\label{P4_LA} &\mathcal{L}(t,\bm{R}_{\mathrm{C}},\bm{R}_{\mathrm{S}},\bm{Z}_B,\mu_R,\mu_P,\bm{Z}_{\mathrm{C}},\bm{Z}_{\mathrm{S}})  \\[-1mm]
	=&t+\mathrm{tr}(\bm{Z}_{\mathrm{C}}\bm{R}_{\mathrm{C}})+\mathrm{tr}(\bm{Z}_{\mathrm{S}}\bm{R}_{\mathrm{S}})-\mu_P(\mathrm{tr}(\bm{R}_{\mathrm{C}}+\bm{R}_{\mathrm{S}})-P) \nonumber\\[-1mm]
	+&\mu_R(\bm{h}^H\bm{R}_{\mathrm{C}}\bm{h}-\gamma(\sigma_{\mathrm{C}}^2 +\bm{h}^H\bm{R}_{\mathrm{S}}\bm{h}))+ \mathrm{tr}(\bm{Z}_B\bm{B}(t,\bm{R}_{\mathrm{C}},\bm{R}_{\mathrm{S}})),\nonumber
\end{align}
where $\mathrm{tr}\left(\bm{Z}_B\bm{B}\left(t,\bm{R}_{\mathrm{C}},\bm{R}_{\mathrm{S}}\right)\right)=-z_1t+\mathrm{tr}\left(\bm{D}(\bm{R}_{\mathrm{C}}+ \bm{R}_{\mathrm{S}})\right) $ with $\bm{D}=z_1\bm{A}_1+z_2\bm{A}^H_2+z_2^*\bm{A}_2+z_3\bm{A}_3$. Then, the Lagrange dual function of Problem (P2-R) is given by
\vspace{-1.5mm}
\begin{align}
	&L_{\mathrm{D}}( \bm{Z}_B, \mu_R,\mu_P,\bm{Z}_{\mathrm{C}},\bm{Z}_{\mathrm{S}}) \nonumber \\
	=&\max\limits_{t,\bm{R}_{\mathrm{C}},\bm{R}_{\mathrm{S}}} \ \mathcal{L}(t,\bm{R}_{\mathrm{C}},\bm{R}_{\mathrm{S}},\bm{Z}_B,\mu_R,\mu_P,\bm{Z}_{\mathrm{C}},\bm{Z}_{\mathrm{S}}).
\end{align}
The dual problem is thus expressed as
\vspace{-1.5mm}\begin{equation}
	\min\limits_{\bm{Z}_B,\mu_R,\mu_P,\bm{Z}_{\mathrm{C}},\bm{Z}_{\mathrm{S}}}\ L_{\mathrm{D}}( \bm{Z}_B,\mu_R,\mu_P,\bm{Z}_{\mathrm{C}},\bm{Z}_{\mathrm{S}}).
	\vspace{-1.5mm}\end{equation}

Denote the optimal primal and dual solutions for Problem (P2-R) as $t^\star$, $\bm{R}_{\mathrm{C}}^\star$, $\bm{R}_{\mathrm{S}}^\star$, and $\bm{Z}_B^\star=[z_1^\star,z_2^\star;z_2^{\star^*},z_3^\star] \succeq \bm{0}$, $\mu_R^\star\geq0$, $\mu_P^\star\geq0$, $\bm{Z}_{\mathrm{C}}^\star \succeq \bm{0}$, $\bm{Z}_{\mathrm{S}}^\star \succeq \bm{0}$, respectively. Since Problem (P2-R) is convex and satisfies Slater's condition, strong duality holds. Therefore, the optimal primal and dual solutions satisfy the Karush-Kuhn-Tucker (KKT) conditions \cite{CVX}, including the feasibility constraints and the following conditions:
\vspace{-1.5mm}
\begin{align} \mathrm{tr}\left(\bm{Z}_B^\star\bm{B}\left(t^\star,\bm{R}^\star_{\mathrm{C}},\bm{R}^\star_{\mathrm{S}}\right) \right)=&0 \label{KKT1}\\[-1mm]		
	\mu_R^\star(\bm{h}^H \bm{R}_{\mathrm{C}}^\star\bm{h}- {\gamma} \left(\sigma_{\mathrm{C}}^2 + \bm{h}^H \bm{R}_{\mathrm{S}}^\star\bm{h}\right))=&0\label{KKTrate}\\[-1mm]	
	\mu_P^\star(\mathrm{tr}\left( \bm{R}_{\mathrm{C}}^\star+\bm{R}_{\mathrm{S}}^\star\right)-P)=&0 \label{KKT_power}\\[-1mm]
	\mathrm{tr}(\bm{Z}_{\mathrm{C}}^\star \bm{R}_{\mathrm{C}}^\star)=&0 \label{KKT2}\\[-1mm]	
	\mathrm{tr}(\bm{Z}_{\mathrm{S}}^\star \bm{R}_{\mathrm{S}}^\star)=&0 \label{KKT3}\\[-1mm]	
	\frac{\partial \mathcal{L}(t^\star,\bm{R}_{\mathrm{C}}^\star,\bm{R}_{\mathrm{S}}^\star,\bm{Z}_B^\star,\mu_R^\star,\mu_P^\star,\bm{Z}_{\mathrm{C}}^\star,\bm{Z}_{\mathrm{S}}^\star)  }{\partial t^\star}=1-z_1^\star=&0 \label{derivative_1}\\[-0.5mm]	
	\frac{\partial \mathcal{L}(t^\star,\bm{R}_{\mathrm{C}}^\star,\bm{R}_{\mathrm{S}}^\star,\bm{Z}_B^\star,\mu_R^\star,\mu_P^\star,\bm{Z}_{\mathrm{C}}^\star,\bm{Z}_{\mathrm{S}}^\star)  }{\partial \bm{R}_{\mathrm{C}}^\star}=&\bm{0} \label{derivative_2}\\[-1mm]	
	\frac{\partial \mathcal{L}(t^\star,\bm{R}_{\mathrm{C}}^\star,\bm{R}_{\mathrm{S}}^\star,\bm{Z}_B^\star,\mu_R^\star,\mu_P^\star,\bm{Z}_{\mathrm{C}}^\star,\bm{Z}_{\mathrm{S}}^\star) }{\partial \bm{R}_{\mathrm{S}}^\star}=&\bm{0}. \label{derivative_3}
\end{align}

Note that from (\ref{derivative_1}), we have $z^\star_1=1$, and consequently $\bm{Z}^\star_B \neq \bm{0}$. From (\ref{KKT1}), we have $|\bm{Z}_B^\star|=z^\star_1z^\star_3-|z^\star_2|^2=0$. Thus, we have $z^\star_3=|z^\star_2|^2$ and $\bm{Z}_B^\star=[1,z_2^\star;z_2^{\star^*},|z_2^\star|^2]$. $\bm{D}^\star$ can be consequently expressed as $\bm{D}^\star=\bm{A}_1+z^\star_2\bm{A}^H_2+z_2^{\star^*}\bm{A}_2+|z^\star_2|^2\bm{A}_3=\int_{-\pi}^\pi \bar{\bm{D}}(\theta) p_\Theta(\theta)d\theta$, where $\bar{\bm{D}}(\theta)$ is given by
\vspace{-2mm}
\begin{align}\label{matrixD}
	&\bar{\bm{D}}(\theta)=[\bm{a}(\phi,\theta), \dot{\bm{a}}(\phi,\theta)]\times\nonumber\\[-1mm]
	&\left[ \begin{array}{ll}
		\|\dot{\bm{b}}(\phi,\theta)\|^2+ |z^\star_2|^2N_r & z^\star_2N_r\\[-0.5mm]
		z^{\star^*}_2 N_r & N_r
	\end{array}\right]
	\left[ \begin{array}{l }
		\bm{a}^H(\phi,\theta)    \\[-0.5mm]
		\dot{\bm{a}}^H(\phi,\theta)
	\end{array}\right].
\end{align}
Note that $\bm{D}^\star$ can be shown to be a positive semi-definite matrix. We express the eigenvalue decomposition (EVD) of $\bm{D}^\star$ as $\bm{D}^\star=\bm{Q}_D\bm{\Lambda}_D\bm{Q}_D^H$, where $\bm{\Lambda}_D=\mathrm{diag}\{d_1,...,d_{N_t}\}$ with $d_1\geq d_2\geq ...\geq d_{N_t}\geq 0$; $\bm{Q}_D=[\bm{q}_1,...,\bm{q}_{N_t}]$ is a unitary matrix with $\bm{Q}_D\bm{Q}_D^H=\bm{Q}_D^H\bm{Q}_D=\bm{I}_{N_t}$. Let $E_n$ with $0<E_n<N_t$ denote the number of the largest eigenvalue of $\bm{D}^\star$, which indicates $d_1=...=d_{E_n}>d_{E_n+1}$. Define $\bm{V} \in \mathbb{C}^{N_t\times E_n}\triangleq[\bm{q}_1,...,\bm{q}_{E_n}]$ as the collection of eigenvectors corresponding to the $E_n$ largest eigenvalues.

Based on (\ref{derivative_2}) and (\ref{derivative_3}), $\bm{Z}^\star_{\mathrm{C}}$ and $\bm{Z}^\star_{\mathrm{S}}$ are given by
\vspace{-1.5mm}
\begin{equation}
	\bm{Z}^\star_{\mathrm{C}} = \mu^\star_P\bm{I}_{N_t}-(\bm{D}^\star + \mu^\star_R\bm{h}\bm{h}^H), \label{Zc}
\vspace{-1.5mm}	\end{equation}
\begin{equation}
	\bm{Z}^\star_{\mathrm{S}} = \mu^\star_P\bm{I}_{N_t}-(\bm{D}^\star - \mu^\star_R\gamma\bm{h}\bm{h}^H).\label{Zs}
	\vspace{-1.5mm}\end{equation}
Since $\bm{Z}^\star_{\mathrm{C}}\succeq \bm{0}$ and $\bm{Z}^\star_{\mathrm{S}}\succeq\bm{0}$ hold, we have $ \mu^\star_P \geq \lambda_{1}( \bm{D}^\star+\mu^\star_R\bm{h}\bm{h}^H)$ and $\mu^\star_P \!\geq\!\lambda_{1}( \bm{D}^\star \!-\! \mu^\star_R \gamma \bm{h}\bm{h}^H )$, where $ \lambda_{i}\left(\cdot\right)$ represents the $i$-th largest eigenvalue. Moreover, to satisfy the communication rate constraint in (\ref{P2R_C2}), we have $\bm{R}_{\mathrm{C}}^\star\neq \bm{0}$. Thus, based on (\ref{KKT2}), we have $|\bm{Z}_{\mathrm{C}}^\star|=0$, which implies
\vspace{-1.5mm}\begin{equation} \label{muP1}
	\mu^\star_P \!=\!\lambda_{1}\left( \bm{D}^\star +  \mu^\star_R\bm{h}\bm{h}^H \right)\!\geq\!\lambda_{1}\!\left( \bm{D}^\star \!-\! \mu^\star_R \gamma \bm{h}\bm{h}^H \right).
	\vspace{-1.5mm}\end{equation}
Then, we analyze the ranks of $\bm{R}^\star_{\mathrm{C}}$ and $\bm{R}^\star_{\mathrm{S}}$ in three cases.

{\bf{Case I}}: $\mu^\star_R=0$. Based on (\ref{KKTrate}), $\mu^\star_R=0$ indicates $\bm{h}^H \bm{R}^\star_{\mathrm{C}}\bm{h}\geq{\gamma} (\sigma_{\mathrm{C}}^2 + \bm{h}^H \bm{R}^\star_{\mathrm{S}}\bm{h})$, i.e., the communication rate constraint is generally inactive, which can correspond to a low rate target $\bar{R}$. With $\bm{Z}^\star_B=[1,z_2^\star;z_2^{\star^*},|z_2^\star|^2]$ and the resulting $\bm{D}^\star$, $\bm{R}^\star_{\mathrm{C}}$ and $\bm{R}^\star_{\mathrm{S}}$ can be obtained by solving (P2-R-I):
\vspace{-1.5mm}\begin{equation}		
	\mbox{(P2-R-I)}\ \underset{ \bm{R}_{\mathrm{C}}\succeq \bm{0},\bm{R}_{\mathrm{S}}\succeq \bm{0}:\mathrm{tr}(\bm{R}_{\mathrm{C}}  +  \bm{R}_{\mathrm{S}})\leq P}{\max}\ \mathrm{tr}(\bm{D}^\star (\bm{R}_{\mathrm{C}}+\bm{R}_{\mathrm{S}})).
	\vspace{-1.5mm}\end{equation}
Based on Proposition 2 in \cite{ISIT}, the optimal solution to (P2-R-I) can be shown to be $\bm{R}^\star_{\mathrm{C}}+\bm{R}^\star_{\mathrm{S}}=P\bm{q}_1\bm{q}_1^H$. Thus, $\mathrm{rank}(\bm{R}^\star_{\mathrm{C}})= 1$ and  $\mathrm{rank}(\bm{R}^\star_{\mathrm{S}})\leq 1$ hold. This indicates that the SDR is \emph{tight}, and the optimal beamforming structure is $\bm{w}=\sqrt{\xi P}\bm{q}_1$ and $\bm{S}=\bm{s}_1=\sqrt{(1-\xi)P}\bm{q}_1$, with $\xi\in [0,1]$ denoting the power allocation coefficient. By noting that $\xi$ does not affect the periodic PCRB, the optimal $\xi$ should be $1$ to maximize the rate. Hence, the optimal solution to (P1) in Case I is $\bm{w}_{\mathrm{I}}^\star=\sqrt{P}\bm{q}_1,\ \bm{S}_{\mathrm{I}}^\star=\bm{0}$, i.e., \emph{no} dedicated sensing beam is needed.

{\bf{Case II}}: $\mu^\star_R>0$, $\lambda_{1}(\bm{D}^\star+\mu^\star_R\bm{h}\bm{h}^H )>\lambda_{1}(\bm{D}^\star-\mu^\star_R \gamma \bm{h}\bm{h}^H)$. Based on (\ref{Zs}) and (\ref{muP1}), we have $\bm{Z}^\star_{\mathrm{S}}\succ \bm{0}$ and $\mathrm{rank}(\bm{Z}^\star_{\mathrm{S}}) = N_t$, which yield $\bm{R}^\star_{\mathrm{S}}=\bm{0}$ due to (\ref{KKT3}). Namely, \emph{no} dedicated sensing beam is needed. Moreover, since $\bm{Z}^\star_{\mathrm{C}}=\bm{Z}^\star_{\mathrm{S}}-\mu^\star_R(1+ {\gamma})\bm{h}\bm{h}^H $ and $\mathrm{rank}(\bm{Z}^\star_{\mathrm{S}}) = N_t$, we have $\mathrm{rank}(\bm{Z}^\star_{\mathrm{C}})\geq N_t-1$. By further noting $|\bm{Z}_{\mathrm{C}}^\star|=0$, we have $\mathrm{rank}(\bm{Z}^\star_{\mathrm{C}})= N_t-1$ and consequently $\mathrm{rank}(\bm{R}^\star_{\mathrm{C}}) = 1$ due to (\ref{KKT2}). Namely, the SDR is \emph{tight}. The optimal $\bm{R}^\star_{\mathrm{C}}$ can be obtained by solving (P2-R-II):
\vspace{-1.5mm}\begin{equation}		\mbox{(P2-R-II)}\quad\mathop{\mathrm{max}}_{\bm{R}_{\mathrm{C}}\succeq \bm{0}:\mathrm{tr}(\bm{R}_{\mathrm{C}})\leq P}\quad\mathrm{tr}((\bm{D}^\star+\mu^\star_R\bm{h\bm{h}}^H)\bm{R}_{\mathrm{C}}).
	\vspace{-1.5mm}\end{equation}
(P2-R-II) is similar to (P2-R-I), for which the optimal solution is  $\bm{R}^\star_{\mathrm{C}}=P\bm{\eta}_1\bm{\eta}_1^H$ with $\bm{\eta}_1 \in \mathbb{C}^{N_t\times 1}$ being the eigenvector corresponding to the largest eigenvalue of $\bm{D}^\star+\mu^\star_R\bm{h h}^H$. The optimal solution to (P1) in Case II is thus given by $\bm{w}_{\mathrm{II}}^\star=\sqrt{P}\bm{\eta}_1,\ \bm{S}_{\mathrm{II}}^\star=\bm{0}$. Note that this can correspond to the case where $\gamma$ and consequently the rate target $\bar{R}$ are very large.

{\bf{Case III}}: \emph{$\mu^\star_R>0$, $\lambda_{1}(\bm{D}^\star + \mu^\star_R\bm{h}\bm{h}^H)\!=\!\lambda_{1}(\bm{D}^\star-\mu^\star_R \gamma \bm{h}\bm{h}^H)$.} In this case, we have the following proposition.
\begin{proposition}\label{prop_case3}
	In Case III, the optimal $\bm{R}^\star_{\mathrm{C}}$ and $\bm{R}^\star_{\mathrm{S}}$ to (P2-R) can be expressed as $	\bm{R}^\star_{\mathrm{C}}=\beta_{\mathrm{C}}\bm{f}\bm{f}^H+\sum_{n=1}^{E_n}\beta_n\bm{q}_n\bm{q}_n^H$ and $\bm{R}^\star_{\mathrm{S}}= \sum_{n=1}^{E_n}\tau_n\bm{q}_n\bm{q}_n^H$, where $\beta_{\mathrm{C}}\geq 0$; $\bm{f}^H\bm{V}=\bm{0}$, $\bm{f}^H\bm{h}\neq0$, and $\|\bm{f}\|^2=1$; $\beta_n\geq 0,\tau_n\geq 0,n=1,...,E_n$; and $\beta_{\mathrm{C}}+\sum_{n=1}^{E_n}(\beta_n+\tau_n)=P$. A new optimal solution can be constructed as $\bar{\bm{R}}^\star_{\mathrm{C}}=\beta_{\mathrm{C}}\bm{f}\bm{f}^H$ and $\bar{\bm{R}}^\star_{\mathrm{S}}= (\sum_{n=1}^{E_n}(\beta_n+\tau_n)) \bm{q}_1\bm{q}_1^H$.
\end{proposition}
\begin{IEEEproof}
	Please refer to Appendix B.
\end{IEEEproof}
Thus, in Case III, the SDR is still \emph{tight}, and an optimal solution to (P1) is $\bm{w}_{\mathrm{III}}^\star\!=\!\sqrt{\beta_{\mathrm{C}}}\bm{f}$, $\bm{S}_{\mathrm{III}}^\star\!=\!\bm{s}_{1,\mathrm{III}}^\star\!=\!\sqrt{\sum_{n=1}^{E_n}(\beta_n\!+\!\tau_n)}\bm{q}_1$, i.e., \emph{one} dedicated sensing beam is needed.
\subsection{Summary of the Optimal Solution to Problem (P1)}
To summarize, the SDR is always \emph{tight}, and the optimal solution to (P1) can be found by solving (P2-R). Moreover, \emph{no} dedicated sensing beam is needed in Cases I and II where the rate target $\bar{R}$ is low or high, and only \emph{one} dedicated sensing beam is needed in Case III with moderate rate target $\bar{R}$.
\section{Numerical Results}\label{sec_num}
In this section, we provide numerical results to evaluate the performance of our proposed design. We set $N_t=4\times 4$, $N_r=4\times 5$, $L=25$, $P=10$ dBm, $\beta_0=-30$ dB, $\sigma_{\mathrm{C}}^2=\sigma_{\mathrm{S}}^2=-90$ dBm, $d=\frac{\lambda}{2}$, $r=100$ m, $h_{\mathrm{B}}=10$ m, and $\frac{PL|\alpha|^2}{\sigma_{\mathrm{S}}^2}=-8$ dB. The communication user is located at a height of $h_{\mathrm{U}}=1$ m and an azimuth angle of $\theta_{\mathrm{U}}=0.85$, with a distance $r_{\mathrm{U}}=600$ m from the BS. We consider an LoS channel from the BS to the communication user under the same model as in Section \ref{sec_sensing}. For the PDF of $\theta$, we consider the von-Mises mixture model in Remark 1 with $K=2$; $\theta_1=-1.2$, $\theta_2=-0.6 $; $\kappa_1 =300$, $\kappa_2 =80$; $p_1=0.54$, $p_2=0.46$. For comparison, we consider the following benchmark schemes.
\begin{itemize}
	\item {\bf{Benchmark Scheme 1: Sensing-oriented beamforming design.}} In this scheme, the sensing beamforming matrix $\bm{S}$ is designed to minimize the periodic PCRB under the constraint $\mathrm{tr}(\bm{SS}^H)\leq P$ without the rate constraint, for which the optimal solution is $\bm{S}^\star=\bm{s}_1^\star=\sqrt{P}\bm{q}_1$ \cite{JSAC_MIMO}.
	\item {\bf{Benchmark Scheme 2: Dual-functional beamforming design.}} In this scheme, we do not introduce dedicated sensing beam and only use $\bm{w}$ as a dual-functional beamforming vector, which is designed by solving (P1) with an additional constraint $\bm{S}=\bm{0}$ via SDR.
	\item {\bf{Benchmark Scheme 3: Most-probable angle based beamforming design.}} In this scheme, we design $\bm{w}$ and $\bm{S}$ to minimize the periodic CRB corresponding to the angle with the highest probability, i.e., $\theta_{\max}= \arg\max p_\Theta(\theta)$, under the constraints in (P1), for which the optimal solution can be obtained similarly as that in this paper.
\end{itemize}
\begin{figure} [t]
	\centering
	\includegraphics[width=3in]{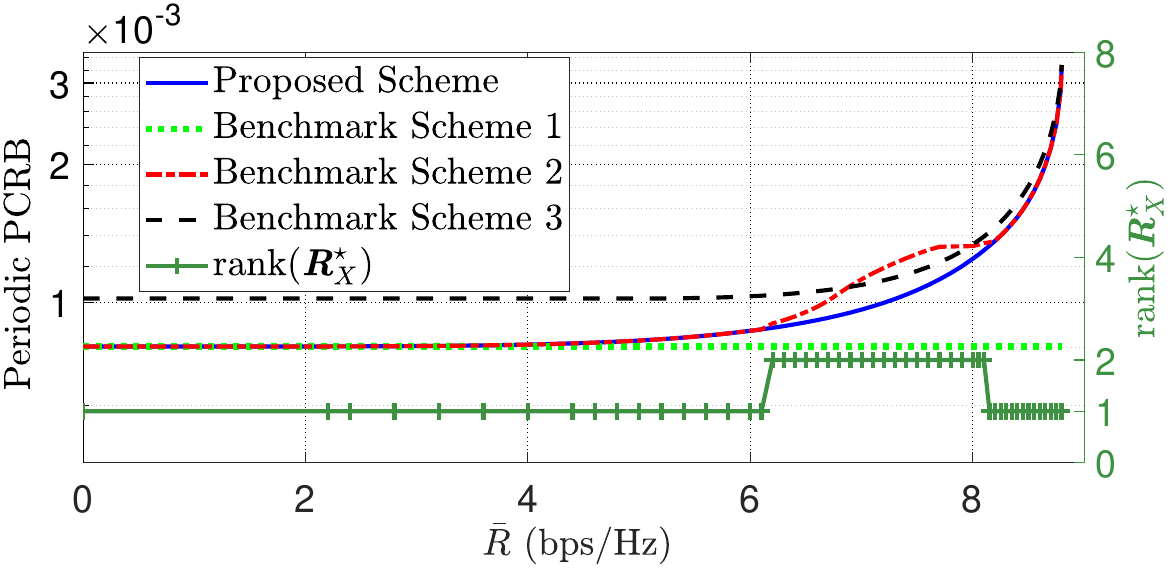}
	\vspace{-4mm}
	\caption{Sensing periodic PCRB versus communication rate target.}\label{Fig_PCRB}
	\vspace{-7.5mm}	\end{figure}

In Fig. \ref{Fig_PCRB}, we show the periodic PCRB of the sensing MCE versus the communication rate target for different designs. We also illustrate the rank of the transmit covariance matrix for the proposed scheme. It is observed that our proposed scheme requires a dedicated sensing beam in the moderate rate regime, and only needs a communication beam in the low or high rate regime, which is consistent with our analytical results in Section \ref{sec_solution}. Moreover, the proposed scheme outperforms Benchmark Scheme 2, since one dual-functional beam may not be sufficient for achieving satisfactory probability-dependent power focusing over the continuous angle range $[-\pi,\pi)$. The proposed scheme also outperforms Benchmark Scheme 3 since the periodic PCRB can fully capture the effect of prior PDF on the sensing performance and lead to more appropriate beamforming design. Finally, it is observed that Benchmark Scheme 1 achieves the lowest PCRB since it only considers the sensing performance, while it can only achieve a very low communication rate of $1.93032$ bps/Hz.
\vspace{-1mm}
\section{Conclusions}
\vspace{-1mm}
This paper investigated the transmit beamforming optimization in an ISAC system where the unknown and random angle parameter of a target needs to be sensed. We first derived the periodic PCRB of the sensing MCE as the sensing performance metric. Then, we jointly optimized the communication and sensing beamforming designs to minimize the periodic PCRB under a minimum communication rate constraint. We derived the optimal solution to this non-convex problem and analytically proved that at most one dedicated sensing beam is needed. Numerical results verified our analysis and showed the effectiveness of our proposed design.




\newpage
\bibliographystyle{IEEEtran}
\bibliography{reference}

\newpage
\appendices
\section{Proof of Proposition \ref{prop_PCRB}}\label{proof_PCRB}
Define $\bm{X}=[\bm{x}_1,...,\bm{x}_L] \in \mathbb{C}^{N_t\times L}$. Let $f(\bm{Y}|\boldsymbol{\zeta})$ denote the conditional PDF of $\bm{Y}$ given $\bm{\zeta}$ and $p_Z(\bm{\zeta})$ denote the marginal distribution of $\boldsymbol{\zeta}$. The joint distribution of the observation $\bm{Y}$ and unknown parameter $\bm{\zeta}$ is given by
\begin{equation}\label{jointpdf}
	f(\bm{Y},\bm{\zeta})=f(\bm{Y}|\bm{\zeta})p_Z(\bm{\zeta}).
\end{equation}

Similar to the conventional PCRB of MSE, the PCRB of MCE is based on the periodic posterior FIM \cite{periodic_PCRB}, which is derived from periodic extended functions of the PDFs. Since $\theta$ is the only periodic parameter in $\bm{\zeta}$, we define an auxiliary vector $\bm{\nu}(\theta)=[- 2\pi\lfloor\frac{\theta+\pi}{2\pi}\rfloor,0,0]^T$, with $\theta \in \mathbb{R}$. The $2\pi$-periodic extended functions are given by $\hat{p}_\Theta(\theta)=  p_\Theta(\theta- 2\pi\lfloor\frac{\theta+\pi}{2\pi}\rfloor)$,  $\hat{p}_Z(\bm{\zeta})= p_Z(\bm{\zeta}+\bm{\nu}(\theta))$, $\hat{f}(\bm{Y}|\boldsymbol{\zeta})= f(\bm{Y}|\bm{\zeta}+ \bm{\nu}(\theta))$, and $\hat{f}(\bm{Y},\bm{\zeta})=\hat{f}(\bm{Y}|\bm{\zeta})\hat{p}_Z(\bm{\zeta})$, with $\theta \in \mathbb{R}$. Note that these periodic extensions are defined for ease of derivation and are not PDFs. Based on the extended function $\hat{f}(\bm{Y},\bm{\zeta})$, the periodic posterior FIM for estimating $\boldsymbol{\zeta}$ is given below and contains two parts \cite{periodic_PCRB}:
\begin{align}\label{FIM}
	\bm{F}  & =\mathbb{E}_{\bm{Y},\boldsymbol{\zeta} }\left[\frac{\partial \ln(\hat{f}(\bm{Y},\boldsymbol{\zeta}))}{\partial \boldsymbol{\zeta} }\left(\frac{\partial \ln(\hat{f}(\bm{Y},\boldsymbol{\zeta}))}{\partial \boldsymbol{\zeta} }\right)^H\right] \nonumber \\
	&=\bm{F}_\mathrm{O} +\bm{F}_\mathrm{P},
\end{align}
where $\bm{F}_\mathrm{O}=\mathbb{E}_{\bm{Y},\boldsymbol{\zeta} }\big[\frac{\partial \ln(\hat{f}(\bm{Y}|\boldsymbol{\zeta}))}{\partial \boldsymbol{\zeta} }\big(\frac{\partial \ln(\hat{f}(\bm{Y}|\boldsymbol{\zeta}))}{\partial \boldsymbol{\zeta} }\big)^H\big]$ and $\bm{F}_\mathrm{P}=\mathbb{E}_{\boldsymbol{\zeta} }\big[\frac{\partial \ln(\hat{p}_Z(\boldsymbol{\zeta} ))}{\partial \boldsymbol{\zeta} }\big(\frac{\partial \ln(\hat{p}_Z(\boldsymbol{\zeta} ))}{\partial \boldsymbol{\zeta} }\big)^H\big]$ represent the FIMs from observation and prior information, respectively.

For $\bm{F}_\mathrm{O}$, based on the expression of $\bm{Y}$, $\ln(f(\bm{Y}|\boldsymbol{\zeta}))$ is expressed as \cite{MIMO_radar}
\begin{align}
	\ln(f(\bm{Y}|\boldsymbol{\zeta}))=&\frac{2}{\sigma_\mathrm{S}^2}\mathfrak{Re}\{ \mathrm{tr}(\bm{X}^H\bm{G}^H(\theta)\bm{Y })\}-N_rL\ln(\pi\sigma_\mathrm{S}^2)\nonumber\\
&-\frac{\|\bm{Y }\|_F^2 + \|\bm{G}(\theta)\bm{X}\|_F^2 }{\sigma_\mathrm{S}^2},\quad \theta\in [-\pi,\pi).
\end{align}
Due to  the periodicity of trigonometric functions, $\bm{G}(\theta- 2\pi\lfloor\frac{\theta+\pi}{2\pi}\rfloor)=\bm{G}(\theta)$ holds for $\theta \in \mathbb{R}$. Thus, $\ln(\hat{f}(\bm{Y}|\boldsymbol{\zeta}))$ can be expressed as
\begin{align} \label{observation}
	\ln(\hat{f}(\bm{Y}|\boldsymbol{\zeta})) =&\frac{2}{\sigma_\mathrm{S}^2}\mathfrak{Re}\{\mathrm{tr}(\bm{X}^H\bm{G}^H(\theta)\bm{Y})\}-N_rL\ln(\pi\sigma_\mathrm{S}^2)\nonumber\\
&-\frac{\|\bm{Y}\|_F^2+\|\bm{G}(\theta)\bm{X}\|_F^2}{\sigma_\mathrm{S}^2},\quad \theta\in\mathbb{R}.
\end{align}
Based on (\ref{observation}), $\bm{F}_\mathrm{O}$ is a function of the derivative of $\bm{G}(\theta)$ given by $\dot{\bm{G}}(\theta)=\alpha(\dot{\bm{b}}(\phi,\theta)\bm{a}(\phi,\theta)^H+\bm{b}(\phi,\theta)\dot{\bm{a}}(\phi,\theta)^H)$. Specifically, $\dot{\bm{a}}(\phi,\theta)$ and $\dot{\bm{b}}(\phi,\theta)$ are the  derivatives of $\bm{a}(\phi,\theta)$ and $\bm{b}(\phi,\theta)$, respectively, which are given by
\begin{align}  \label{dota}
	\dot{ {a}}_n(\phi,\theta)=&j\Delta \Big[ ( T_y-2 ((n-1)\mod{T_y})-1)\cos\theta   \nonumber  \\[-0.5mm]
&-     \Big(T_x-2 \Big\lfloor\frac{n-1}{T_y} \Big\rfloor-1 \Big)\sin\theta\Big]{a}_n(\phi,\theta),
\end{align}
\begin{align}\label{dotb}
	\dot{ {b}}_m(\phi,\theta)=& j\Delta\Big[ (R_y-2 ((m-1)\mod{R_y})-1)\cos\theta \nonumber  \\[-0.5mm]
&-   \Big(R_x-2 \Big\lfloor\frac{m-1}{R_y} \Big\rfloor-1 \Big)\sin\theta\Big]b_m(\phi,\theta).
\end{align}
Note that $\boldsymbol{a}^H(\phi,\theta)\dot{\bm{a}}(\phi,\theta)=0$ and $\boldsymbol{b}^H(\phi,\theta)\dot{\bm{b}}(\phi,\theta)=0$ hold. From (\ref{observation})-(\ref{dotb}), $\bm{F}_\mathrm{O}$ can be expressed as
\begin{align}
	\bm{F}_\mathrm{O}
	= \left[
	\begin{array}{ll}
		J_{\theta\theta}            & \bm{J}_{\theta\alpha}     \\
		\bm{J}_{\theta\alpha}^H & \bm{J}_{\alpha\alpha}
	\end{array}
	\right].
\end{align}
In $\bm{F}_\mathrm{O}$, $J_{\theta\theta}$, $\bm{J}_{\theta\alpha}$, and $\bm{J}_{\alpha\alpha}$  are given by \cite{MIMO_radar,shen}
\begin{align} 		
	J_{\theta\theta}&=\frac{2|\alpha|^2L}{\sigma_\mathrm{S}^2}\mathrm{tr}(\bm{A}_1\bm{R}_X),  \\
	\bm{J}_{\theta \alpha}&= \frac{2L  }{\sigma_\mathrm{S}^2}\mathfrak{Re}\{\mathrm{tr}(\bm{A}_2\bm{R}_X)\alpha[1,j]\},\\
	\bm{J}_{\alpha\alpha}&= \frac{2L  }{\sigma_\mathrm{S}^2}\mathrm{tr}(\bm{A}_3\bm{R}_X)  \bm{I}_{2},
\end{align}
where
\begin{align} 	
	\bm{A}_1=&\int_{-\pi}^{\pi}\Big(\|\dot{\bm{b}}(\phi,\theta)\|^2\bm{a}(\phi,\theta)\bm{a}^H(\phi,\theta)\nonumber\\
	&+N_r\dot{\bm{a}}(\phi,\theta)\dot{\bm{a}}^H(\phi,\theta)\Big) p_\Theta(\theta)d\theta,\\
	\bm{A}_2=&N_r\int_{-\pi}^{\pi} \dot{\bm{a}}(\phi,\theta)\bm{a}^H(\phi,\theta) p_\Theta(\theta)d\theta,\\
	\bm{A}_3=&N_r\int_{-\pi}^{\pi} \bm{a}(\phi,\theta)\bm{a}^H(\phi,\theta) p_\Theta(\theta)d\theta.
\end{align}

On the other hand, for $\bm{F}_\mathrm{P}$, since $\alpha_\mathrm{R}$ and $\alpha_\mathrm{I}$ are deterministic parameters, only the entry   $[\bm{F}_\mathrm{P}]_{1,1}=\mathbb{E}_\theta\big[\big(\frac{\partial \ln(\hat{p}_\Theta(\theta))}{\partial \theta}\big)^2\big]=\int_{-\pi}^{\pi}\big(\frac{\partial \ln(\hat{p}_\Theta(\theta))}{\partial \theta}\big)^2p_\Theta(\theta)d\theta\geq 0$ in the first column and the first row can be non-zero.

Therefore, the periodic posterior FIM for $\boldsymbol{\zeta}$ is given by
\begin{align}
	\boldsymbol{F} \!=\!\bm{F}_\mathrm{O} \!+\! \bm{F}_\mathrm{P}\!=\!\left[
	\begin{array}{cc}
		J_{\theta\theta}\!+\!\mathbb{E}_\theta\left[\left(\frac{\partial \ln(\hat{p}_\Theta(\theta))}{\partial \theta}\right)^2\right]          & \bm{J}_{\theta\alpha}     \\
		\bm{J}_{\theta\alpha}^H & \bm{J}_{\alpha\alpha}
	\end{array}
	\right].
\end{align}

Based on the  Schur complement, $[\bm{F}^{-1}]_{1,1}$ is expressed as
\begin{align}
[\bm{F}^{-1}]_{1,1}=&1 \bigg/\left(\mathbb{E}_\theta\left[\left(\frac{\partial \ln(\hat{p}_\Theta(\theta))}{\partial \theta} \right)^2\right] \right.  \nonumber\\
&+\left.\frac{2|\alpha|^2L}{\sigma_\mathrm{S}^2}\!\left(\!\mathrm{tr}( \bm{A}_1\bm{R}_X) \!-\!\frac{ \left| \mathrm{tr}(\bm{A}_2\bm{R}_X)\right|^2}{ \mathrm{tr}(\bm{A}_3\bm{R}_X)}\!\right)\!\right).
\end{align}
Then, the PCRB of the MCE for the desired sensing parameter $\theta$ is given by $\mathrm{PCRB}_{\theta}^\mathrm{P}= 2-2(1+ [\bm{F}^{-1}]_{1,1})^{-\frac{1}{2}}$ \cite{periodic_PCRB}, which is shown in (\ref{PCRB}). Proposition \ref{prop_PCRB} is thus proved.

\section{Proof of Proposition \ref{prop_case3}}\label{proof_case3}
Since $\bm{h}\bm{h}^H $ is a positive semi-definite matrix, we have $\lambda_{1}( \bm{D}^\star + \mu^\star_R\bm{h}\bm{h}^H )\geq d_1$ and $\lambda_{1}( \bm{D}^\star - \mu^\star_R \gamma \bm{h}\bm{h}^H )\leq d_1$ from Weyl's inequality \cite{Matrix}. Thus, Case III implies that $\mu^\star_P = d_1$ and $\bm{h}^H\bm{V}=\bm{0}$. Then, $\bm{Z}^\star_{\mathrm{C}}$ can be expressed as $\bm{Z}^\star_{\mathrm{C}}=(d_{1}\bm{I}_{N_t}-\bm{D}^\star)-\mu^\star_R\bm{h}\bm{h}^H$. Based on this, we have
\begin{align}
	N_t-E_n-1\leq \mathrm{rank}(\bm{Z}^\star_{\mathrm{C}})\leq N_t-E_n,
\end{align}
where the first inequality holds due to $\mathrm{rank}(\bm{Z}^\star_{\mathrm{C}})\geq\mathrm{rank}(d_{1}\bm{I}_{N_t}-\bm{D}^\star)-\mathrm{rank}(\mu_R^\star\bm{h}\bm{h}^H)$ and the second inequality holds due to $\bm{Z}^\star_{\mathrm{C}}\bm{V}=\bm{0}$. On the other hand, $\bm{Z}^\star_{\mathrm{S}}$ can be expressed as $\bm{Z}^\star_{\mathrm{S}}=(d_{1}\bm{I}_{N_t}-\bm{D}^\star)+\mu^\star_R\gamma\bm{h}\bm{h}^H$. From Weyl's inequality \cite{Matrix}, we have $\lambda_i(\bm{Z}^\star_{\mathrm{S}})\geq\lambda_i(d_{1}\bm{I}_{N_t}-\bm{D}^\star)$, which yields $\mathrm{rank}(\bm{Z}^\star_{\mathrm{S}})\geq\mathrm{rank}(d_{1}\bm{I}_{N_t}-\bm{D}^\star)=N_t-E_n$. Moreover, we have $\mathrm{rank}(\bm{Z}^\star_{\mathrm{S}})\leq N_t-E_n$ from $\bm{Z}^\star_{\mathrm{S}}\bm{V}=\bm{0}$. Therefore, we have $\mathrm{rank}(\bm{Z}^\star_{\mathrm{S}})= N_t-E_n$.

Furthermore, if $\mathrm{rank}(\bm{Z}^\star_{\mathrm{C}})= N_t-E_n$, $\bm{V}$ will be the orthogonal basis for the null space of $\bm{Z}^\star_{\mathrm{C}}$. Then, the optimal $\bm{R}^\star_{\mathrm{C}} $ can be expressed as $\bm{R}^\star_{\mathrm{C}}=\sum_{n=1}^{E_n}\beta_n\bm{q}_n\bm{q}_n^H$, with $\beta_n\geq 0,\forall n$. Since $\bm{h}^H\bm{V}=\bm{0}$, we have $\bm{h}^H \bm{R}^\star_{\mathrm{C}}\bm{h} =0$, which violates the communication rate constraint. Hence, we have $\mathrm{rank}(\bm{Z}^\star_{\mathrm{C}})=N_t-E_n-1$. Let $\bm{U} \in \mathbb{C}^{(E_n+1)\times N_t}=[\bm{V}, \bm{f} ]$ denote the orthogonal basis for the null space of $\bm{Z}^\star_{\mathrm{C}}$. The optimal $\bm{R}^\star_{\mathrm{C}} $ can be expressed as
\begin{align}\label{Rc}
	\bm{R}^\star_{\mathrm{C}}=\beta_{\mathrm{C}}\bm{f}\bm{f}^H+\sum_{n=1}^{E_n}\beta_n\bm{q}_n\bm{q}_n^H,
\end{align}
with $\beta_{\mathrm{C}}\geq 0$ and $\beta_n\geq 0,\ n=1,...,E_n$.

Since $\bm{V}$ is the orthogonal basis for the null space of $\bm{Z}^\star_{\mathrm{S}}$, the optimal $\bm{R}^\star_{\mathrm{S}} $ can be expressed as
\begin{align}\label{Rs}
	\bm{R}^\star_{\mathrm{S}}= \sum_{n=1}^{E_n}\tau_n\bm{q}_n\bm{q}_n^H,
\end{align}
with $\tau_n\geq 0,\ n=1,...,E_n$. Specifically, due to (\ref{KKT_power}) and $\mu_P^\star=d_1>0$, we have $\mathrm{tr}(\bm{R}^\star_{\mathrm{C}}+\bm{R}^\star_{\mathrm{S}})-P=0$, which yields $\beta_{\mathrm{C}}+\sum_{n=1}^{E_n}(\beta_n+\tau_n)=P$.

Given optimal dual variables $\mu^\star_R$ and $\bm{Z}^\star_B$, the optimal solutions $\bm{R}^\star_{\mathrm{C}}$ and $\bm{R}^\star_{\mathrm{S}}$ can be obtained by solving the following problem:
\begin{align}
	\mbox{(P2-R-III)} \underset{\scriptstyle \bm{R}_{\mathrm{C}}\succeq \bm{0}, \bm{R}_{\mathrm{S}}\succeq \bm{0}:\atop\mathrm{tr}(\bm{R}_{\mathrm{C}}  +  \bm{R}_{\mathrm{S}})\leq P}{\max}  &  \mathrm{tr}(\bm{D}^\star(\bm{R}_{\mathrm{C}}+\bm{R}_{\mathrm{S}}))\nonumber\\
	&\qquad+\mu^\star_R\bm{h}^H(\bm{R}_{\mathrm{C}} -{\gamma} \bm{R}_{\mathrm{S}})\bm{h}.
\end{align}

Based on the optimal solutions in (\ref{Rc}) and (\ref{Rs}), we construct an alternative solution given by $ \bar{\bm{R}}^\star_{\mathrm{C}}=\beta_{\mathrm{C}}\bm{f}\bm{f}^H$ and $\bar{\bm{R}}^\star_{\mathrm{S}}= (\sum_{n=1}^{E_n}(\beta_n+\tau_n)) \bm{q}_1\bm{q}_1^H$ to Problem (P2-R-III). Note that the optimal value of (P2-R-III) is given by
\begin{align}
	&\mathrm{tr}(\bm{D}^\star(\bm{R}^\star_{\mathrm{C}}+\bm{R}^\star_{\mathrm{S}}))+\mu^\star_R \bm{h}^H( \bm{R}^\star_{\mathrm{C}}- {\gamma}\bm{R}^\star_{\mathrm{S}})\bm{h}\nonumber \\
	=&\sum_{n=1}^{E_n}(\beta_n+\tau_n)d_n+\beta_{\mathrm{C}}\bm{f}^H \bm{D}^\star\bm{f}+\beta_{\mathrm{C}}\mu^\star_R|\bm{f}^H \bm{h}|^2.
\end{align}
With $\bar{\bm{R}}^\star_{\mathrm{C}}$ and $\bar{\bm{R}}^\star_{\mathrm{S}}$, the objective function of (P2-R-III) is
\begin{align}
	&\mathrm{tr}(\bm{D}^\star(\bar{\bm{R}}^\star_{\mathrm{C}}+\bar{\bm{R}}^\star_{\mathrm{S}}))+\mu^\star_R\bm{h}^H (  \bar{\bm{R}}^\star_{\mathrm{C}}- {\gamma}\bar{\bm{R}}^\star_{\mathrm{S}})\bm{h}\nonumber \\
	=&\sum_{n=1}^{E_n}(\beta_n+\tau_n)d_1+\beta_{\mathrm{C}}\bm{f}^H \bm{D}^\star\bm{f}+\beta_{\mathrm{C}}\mu^\star_R|\bm{f}^H \bm{h}|^2.
\end{align}
Since $d_1=...=d_{E_n}$, the objective function of (P2-R-III) achieved by $(\bar{\bm{R}}^\star_{\mathrm{S}},\bar{\bm{R}}^\star_{\mathrm{C}})$ is the same as that achieved  by $( {\bm{R}}^\star_{\mathrm{S}}, {\bm{R}}^\star_{\mathrm{C}})$. Moreover, $(\bar{\bm{R}}^\star_{\mathrm{S}}, \bar{\bm{R}}^\star_{\mathrm{C}})$ is a feasible solution since $ \bar{\bm{R}}^\star_{\mathrm{C}} \succeq \bm{0}$, $ \bar{\bm{R}}^\star_{\mathrm{S}} \succeq \bm{0}$, and $\mathrm{tr}( \bar{\bm{R}}^\star_{\mathrm{S}}+\bar{\bm{R}}^\star_{\mathrm{C}} )=\beta_{\mathrm{C}}+\sum_{n=1}^{E_n}(\beta_n+\tau_n)= P$.

Therefore, $(\bar{\bm{R}}^\star_{\mathrm{S}}, \bar{\bm{R}}^\star_{\mathrm{C}})$ is an optimal solution for Problem (P2-R-III) and Problem (P2-R) in Case III. This thus completes the proof of Proposition \ref{prop_case3}.

\end{document}